\documentclass[twocolumn,nofootinbib,amsmath,amssymb,a4paper]{revtex4}
\usepackage{graphicx}
\usepackage{dcolumn}
\usepackage{bm}
\usepackage{amsmath}
\RequirePackage{xspace}

\def\ra                 {\ensuremath{\rightarrow}\xspace}
\def\CP                {\ensuremath{C\!P}\xspace}
\newcommand{\Nbtoappim}{\mbox{$B^0 \rightarrow a^{\pm}_1(1260)\, \pi^{\mp}  $}}
\newcommand{\Natopipipi}{\mbox{$ a^{\pm}_1(1260) \rightarrow \pi^{\mp}\pi^{\pm}\pi^{\pm}  $}}
\newcommand{\NNbtoappim}{\mbox{$B^0 \rightarrow a^{\pm}_1\, \pi^{\mp}  $}}
\newcommand{\aunobpi}{\mbox{$a_1 \pi$}}
\newcommand{\aunorhopi}{\mbox{$a_1 \rightarrow \rho \pi$}}

\newcommand{\aunoppim}{\mbox{$a_1^+ \pi^-$}}
\newcommand{\aunompip}{\mbox{$a_1^- \pi^+$}}

\newcommand{\aunocpi}{\mbox{$a_1^{q} \pi^{-q}$}}
\newcommand{\aunopi}{\mbox{$a_1^{\pm} \pi^{\mp}$}}
\newcommand{\Nbtoadueppim}{\mbox{$B^0 \rightarrow a^{\pm}_2(1320)\, \pi^{\mp}  $}}
\newcommand{\Nadueppim}{\mbox{$a^{\pm}_2(1320)\, \pi^{\mp}$}}

\def\babar{{\em B}{\footnotesize\em A}{\em B}{\footnotesize\em AR}}
\def\BB{\mbox{$B\overline B\ $}}
\def\pep2{PEP-II}
\def\Acpapi{{\cal A}_{CP}^{\aunobpi}}

\def\dCa1pi{\ensuremath{\Delta C}_{\aunobpi}}
\def\dSa1pi{\ensuremath{\Delta S}_{\aunobpi}}
\def\Ca1pi{\ensuremath{C_{\aunobpi}}}
\def\Sa1pi{\ensuremath{S_{\aunobpi}}}

\def\deltat {\ensuremath{\Delta t}\xspace}
\def\epem  {\ensuremath{e^+e^-}\xspace}
\def\deltamd{\ensuremath{{\rm \Delta}m_d}\xspace}
\def\angular{\mbox{$\cal H$}}
\newcommand{\bflav}{\ensuremath{B_{\rm flav}}}

\newcommand{\beqn}{\begin{eqnarray}}
\newcommand{\eeqn}{\end{eqnarray}}
\newcommand{\ttag}{\ensuremath{t_{\rm tag}}}

\newcommand{\aunob}{\mbox{$a_1$}}
\def\a1pi{\aunob\pi}
\def\Bz      {\ensuremath{B^0}}
\def\Bbar    {\overline{B}{}}
\def\Bzb     {\ensuremath{\Bbar^0}}
 \def\mes{\mbox{$m_{\rm ES}$}}
\def\BzBzb   {\ensuremath{\Bz  \Bzb}}
\def\Bu      {\ensuremath{B^+}}
\def\Bub     {\ensuremath{B^-}}
\def\BpBm    {\ensuremath{\Bu  \Bub}}
\newcommand{\gev}{\mbox{$\textrm{GeV}$}}
\newcommand{\mev}{\mbox{$\textrm{MeV}$}} 
\newcommand{\pvec}{{\bf p}}
\newcommand{\DE}{\ensuremath{\Delta E}}
  \newcommand{\half}{\mbox{${1\over2}$}}
\newcommand{\UfourS}{\mbox{$\Upsilon(4S)$}}

\def\bbar  {\mbox{$\bar{b} $}}
\def\dbar  {\mbox{$\bar{d} $}}
\def\ubaru{\mbox{$\bar{u}  u $}}
\def\qqbar{\mbox{$q \bar{q} $}}

\newcommand{\xf}{\mbox{${\cal F}$}}
\def\KS    {\ensuremath{K^0_{\scriptscriptstyle S}}}
\newcommand\etal{{\it et al.}}
\newcommand{\jprlBase}  [1]     {Phys.\ Rev.\ Lett. \xspace}
\newcommand{\jprl}      [1]    {\jprlBase\ {\bf #1}}
\newcommand{\jprBase}        {Phys.\ Rev.\ \xspace}
\newcommand{\jprd}      [1]  {\jprBase\ D~{\bf #1}}
\newcommand{\plBase}   [1]         {Phys.\ Lett.\xspace}
\newcommand{\plb}      [1]    {\plBase\ B~{\bf #1}}
\newcommand{\nimBaseA}       {Nucl.\ Instr.\ Meth.\ \xspace }
\newcommand{\nima}      [1]  {\nimBaseA~A~{\bf #1}}

\newcommand{\npBase}         {Nucl.\ Phys.\ \xspace}
\newcommand{\npb}       [1]  {\npBase\ B~{\bf #1}}
\newcommand{\jpg}       [1]  {{J.\ Phys.\ {\bf G{\bf #1}}}}
\newcommand{\progtp}    [1]  {{Prog.\ Theor.\ Phys.\ {\bf #1}}}
\newcommand{\jmplBase}  [1]     {Mod.\ Phys.\ Lett.}

\newcommand{\epjBase}  [1]     {Eur.\ Phys.\ J. \xspace}
\newcommand{\epj}      [1]    {\epjBase\ C~{\bf #1}}

\begin{document}

\title{\large  \bf\boldmath  Measurements of \CP-Violating Asymmetries in \Nbtoappim\ Decays}

\author{F. Palombo}
 \email{palombo@mi.infn.it}
\affiliation{%
Dipartimento di Fisica, Universit\`a degli Studi di Milano and INFN, I-20133 Milano, Italy
}

\begin{abstract}
 We present measurements  of \CP-violating asymmetries in the decays 
\Nbtoappim\  with \Natopipipi. The data sample corresponds to $384
\times 10^6$ \BB\ pairs  collected with the \babar\ detector at the \pep2
asymmetric $B$-factory at SLAC. We measure the time- and flavor-integrated charge asymmetry
$\Acpapi = -0.07 \pm 0.07 \pm 0.02$, 
 the mixing-induced \CP violation parameter  $\Sa1pi = 0.37 \pm 0.21\pm 0.07$, the direct \CP violation 
parameter $\Ca1pi = -0.10 \pm 0.15\pm 0.09$, and the parameters $\dCa1pi = 0.26 \pm 0.15\pm 0.07$ 
and $\dSa1pi = -0.14 \pm 0.21 \pm 0.06$. From these measured quantities we extract  the angle 
$\alpha_{\rm eff} = 78.6^{\circ} \pm 7.3^{\circ}$.
\end{abstract}

\maketitle

\section{Introduction}
The angle $\alpha \equiv \arg\left[-V_{td}^{}V_{tb}^{*}/V_{ud}^{}V_{ub}^{*}\right]$
of the unitarity triangle of the Cabibbo-Kobayashi-Maskawa (CKM) quark-mixing
 matrix \cite{CKM} has recently been measured
by the \babar\ and Belle Collaborations
from time-dependent \CP asymmetries in the $\bbar \ra \ubaru \dbar $ dominated 
 $B^0$ decays to $\pi^+\pi^-$ \cite{pipi},
$\rho^{\pm}\pi^{\mp}$ \cite{rhopi}, and $\rho^+ \rho^-$ \cite{rhorho}.
In all these rare $B$ decays  the presence of additional loop (penguin) contributions with a 
different weak phase than the  $\bbar \ra \ubaru \dbar $ tree amplitudes
complicates the extraction of the angle $\alpha$. 
Theoretical uncertainties \cite{Zupan2} and available experimental data samples limit the current  
precision  on this measurement.  Therefore  a new and independent measurement of the 
angle $\alpha$  in another  $B$ decay mode is important to increase the precision 
of the measurement. 

The decays $B^0 \ra \aunopi$ \footnote{For the $a_1(1260)$ meson we  use the short notation
${a_1}$.} proceed dominantly through the
  $\bbar  \ra  \ubaru \dbar $    process in the same way as the
     previously studied modes and can be used  to measure the time-dependent 
\CP asymmetries and extract the angle $\alpha$ \cite{Aleksan}.
 The observation of these $B^0$ decay modes
has been recently reported by the \babar\ collaboration \cite{a1pi}.

\section{Analysis Method  }
\subsection{Strategy in the Measurement of $\alpha$ in the Decays  $\Bz(\Bzb) \ra  a_1^{\pm}\pi^{\mp}$.}
In these proceedings we report the measurements of the \CP-violating asymmetries  in the 
decays \NNbtoappim\ with  $\aunob^{\pm} \ra \pi^{\mp} \pi^{\pm} \pi^{\pm}$~\cite{Vincenzo}.
 These asymmetries  may be  then used to extract the angle $\alpha$. As mentioned in the 
introduction, this 
extraction is complicated by the presence of penguin contributions. We might  overcome 
these complications using  isospin symmetry \cite{GL} or a time-dependent Dalitz plot
 analysis \cite{Snyder}
or approximate SU(3) flavor  symmetry \cite{Gross}. 
The state $a_1^{\pm}\pi^{\mp}$, like  $\rho^{\pm}\pi^{\mp}$,  is not a \CP
 eigenstate and four flavor-charge configurations must be considered
 $(\Bz(\Bzb) \ra  a_1^{\pm}\pi^{\mp}$). Symmetry applications are similar in the 
   $B^0 \ra \rho^{\pm}\pi^{\mp}$ and $B^0 \ra a_1^{\pm}\pi^{\mp}$ decay modes.
A full isospin analysis \cite{GL} requires the precise measurement of the branching fractions 
and asymmetries in the five modes 
$B^0 \ra a_1^+ \pi^- ,\, a_1^- \pi^+ ,\, a_1^0 \pi^0 ,\, B^+ \ra a_1^+ \pi^0 ,\, a_1^0 \pi^+$
and in the five charge conjugate modes.  
Currently only the first two decay modes (and the corresponding 
two charge conjugate modes )  have been studied  experimentally \cite{a1pi}. 
However, even measuring all the ten  branching fractions and the time-dependent \CP asymmetries 
in the three  $B^0$ decay  modes, this isospin method for the extraction of the angle $\alpha$
is not feasible at the present statistics
because of the inaccuracy expected on the measured experimental quantities.

As pointed out in Ref.~\cite{Snyder} the angle $\alpha$ may be 
extracted without ambiguity with a 
time-dependent analysis on the Dalitz plot . This method has been recently  applied to the  
decay $B^0 \ra \pi^+ \pi^- \pi^0$   \cite{Matt}. It could be applied to the decay 
$B^0 \ra \pi^+ \pi^- \pi^0 \pi^0$ with contributions from $a_1^+ \pi^-$, $a_1^- \pi^+$, 
$a_1^0 \pi^0$, $\rho^+ \rho^-$ amplitudes or to the decay  
$B^0 \ra \pi^+ \pi^- \pi^+ \pi^-$  with contributions from $a_1^+ \pi^-$, $a_1^- \pi^+$, 
and $\rho^0 \rho^0$ amplitudes.
Such analyses would be difficult because of the four particles in the final state, 
uncertainties in the  $a_1$ meson parameters and lineshape, the small number of 
signal events and the large expected  background. With current data samples 
this approach seems  impractical. It could  be considered in the next years when more 
data will be available.  

The BaBar analysis  presented in these proceedings follows a quasi-two-body 
approximation. The decays  $\Bz(\Bzb) \ra  a_1^{\pm}\pi^{\mp}$ have been reconstructed 
with  $a_1^{\pm} \ra \pi^{\mp} \pi^{\pm} \pi^{\pm}$ 
(all charged particles in final state). The other sub-decay modes with 
  $a_1^{\pm} \ra \pi^{\pm} \pi^{0} \pi^{0}$ could be used to enhance statistics 
but they have low reconstruction 
efficiency and large background. Details on the  reconstraction
and handling of the \aunob\ meson can be found in Ref.~\cite{a1pi}.
From a time-dependent \CP analysis we extract an effective angle
 $\alpha_{\rm eff}$ which is an approximate measure of the angle $\alpha$ \cite{Charles}. These two
angles coincide in the limit of vanishing  penguin contributions.
Details on this approach
for the decays   $B^0 \ra  a_1^{\pm} \pi^{\mp}$  can be found in Ref.\cite{Zupan}.
Applying flavor SU(3) symmetry one can determine an upper  bound on  
$\Delta \alpha =| \alpha -\alpha_{\rm eff}|$,
using the ratio of \CP-averaged rates involving
SU(3) related decays (in the  axial-vector nonet $1^{++}$): $B^0 \ra a_1^+ K^-$,
$B^0 \ra K^+_1(1270) \pi^-$, $B^0 \ra K^+_1(1400) \pi^-$ or  $B^+ \ra a_1^+ K^0$,
$B^+ \ra K^0_1(1270) \pi^+$, $B^+ \ra K^0_1(1400) \pi^+$.

\subsection{Time-Dependence }
From a candidate \BB\ pair we reconstruct a \Bz\  decaying into the final 
state $f= \aunobpi$ ($B^0_{\aunobpi}$). We also reconstruct the vertex of 
the other $B$ meson ($B^0_{\rm tag}$) and identify its flavor.
The difference $\deltat \equiv t_{\aunobpi} - \ttag$
of the proper decay times of the reconstructed and tag $B$ mesons, 
respectively, is obtained from the measured distance between the $B^0_{\aunobpi}$
and  $B^0_{\rm tag}$ decay vertices and from the boost ($\beta \gamma =0.56$) of 
the \epem system. The \deltat\ distributions are given \cite{Zupan} by:
\beqn
\label{eq:thTime}
  \lefteqn{F^{a_1^{\pm} \pi^{\mp}}_{Q_{\rm tag}}(\deltat) = (1\pm \Acpapi)
           \frac{e^{-\left|\deltat\right|/\tau}}{4\tau} \bigg\{ 1 - Q_{\rm
             tag} \Delta w +}\\
	&&\hspace{1.cm} Q_{\rm tag} (1-2w) 
             \bigg[(\Sa1pi \pm \dSa1pi)\sin(\deltamd\deltat)-\nonumber\\[-0.1cm]
	&&\hspace{1.cm}\phantom{Q_{\rm tag} (1-2w) \bigg[}
		(\Ca1pi\pm \dCa1pi)\cos(\deltamd\deltat)\bigg]\bigg\}\;,\nonumber
\eeqn
where $Q_{\rm tag}= +1(-1)$ when the tagging meson $\Bz_{\rm tag}$
is a $\Bz(\Bzb)$, $\tau$ is the mean 
\Bz\ lifetime, $\deltamd$ is the mass difference between the two $B^0$ 
mass eigenstates, and the mistag parameters $w$ and $\Delta w$ are the
average and difference, respectively, of the probabilities that a true
$\Bz$ is incorrectly tagged as a $\Bzb$ or vice versa.
The time- and flavor-integrated  charge asymmetry $\Acpapi$ measures 
direct \CP violation.
The quantities $\Sa1pi$ and $\Ca1pi$ 
parameterize the mixing-induced \CP violation related to the angle $\alpha$,
and flavor-dependent direct \CP violation, respectively.
The parameter $\dCa1pi$ describes the asymmetry between the rates 
$\Gamma({\Bz} \to{a_1^+\pi^-}) + \Gamma({\Bzb} \to {a_1^-\pi^+})$ and
${\Gamma(\Bz} \to {a_1^-\pi^+}) + \Gamma({\Bzb} \to {a_1^+\pi^-})$, while
$\dSa1pi$ is related to the strong phase difference between
the amplitudes contributing to $\Bz \to \aunopi$ decays. The parameters $\dCa1pi$ 
and $\dSa1pi$ are insensitive to \CP violation.

A measurable angle  $\alpha_{\rm eff}$ can be defined  \cite{Zupan} as:
\beqn
\label{eq:alfaeff}
\alpha_{\rm eff} = \frac{1}{4}
\bigg[\arcsin \bigg(
\frac{\Sa1pi + \dSa1pi }
{\sqrt{1-(\Ca1pi + \dCa1pi)^2}}
\bigg) + \\
	&&\hspace{-4.7cm}\arcsin \bigg(
\frac{\Sa1pi - \dSa1pi }
{\sqrt{1-(\Ca1pi - \dCa1pi)^2}}
\bigg)\bigg]\;\nonumber
\eeqn

To resolve discrete ambiguities in $\alpha_{\rm eff}$, the relative 
strong  phase of the tree amplitudes of the \Bz\ decays to \aunompip\   and   \aunoppim\
 has been assumed  much smaller than $90^\circ$ \cite{Zupan}, as predicted by QCD factorization \cite{QCDF}
and valid to leading order in $1/m_b$ \cite{Mb} . With this assumption 
$\alpha_{\rm eff}$ can be determined   from formula~\ref{eq:alfaeff} up to four-fold 
discrete ambiguity.  

Charge-flavor specific branching fractions can be obtained through the relation \cite{Charles}:
\beqn
\label{eq:thTimee}
{\cal B}_{\aunocpi}(Q_{\rm tag},q) = \frac{1}{2} (1+q \, \Acpapi)(1+\\
  &&\hspace{-4.7cm}  Q_{\rm tag} (\Ca1pi +q \, \dCa1pi))\, {\cal B}^{\pm \mp}_{\aunobpi}\;\nonumber
\eeqn

with $q$  the charge of the $a_1$ meson 
and ${\cal B}^{\pm \mp}_{\aunobpi}$  the measured branching fraction \cite{a1pi} where 
the  final states $a_1^+ \pi^-$ and $a_1^- \pi^+$ are summed and  intial states ($B$ flavors) 
are  averaged. 

\section{The \babar\ Detector and Dataset  }
The data used in this analysis  were collected  
with the \babar\ detector at the PEP-II asymmetric $e^+e^-$ collider. An integrated
luminosity of 349~fb$^{-1}$, corresponding to
384 $\pm$ 4 million \BB\ pairs, was recorded at the $\Upsilon (4S)$ resonance
(``on-resonance'') at a  center-of-mass (CM) energy $\sqrt{s}=10.58~\gev$.
An additional 37~fb$^{-1}$ were taken about 40~MeV below
this energy (``off-resonance'') for the study of continuum background in
which a charm or lighter quark pair is produced
.
A detailed description of the \babar\ detector is given in Ref.~\cite{BABARNIM}.
Track and vertex reconstruction is based on a silicon vertex tracker (SVT) and a drift 
chamber (DCH). Photons are reconstructed in an electromagnetic calorimeter (EMC). The 
internally reflected Cherenkov  light together with the energy loss (dE/dx) in the SVT and DCH
are used for particle identification. Muons are primarily identified by the use of the 
instrumented flux return of the solenoid.

\section{Event Selection and Background Suppression}
Full Monte Carlo (MC) simulations of the signal decay modes, 
continuum, and \BB\ backgrounds   are used to establish 
the event selection criteria.
The MC signal events are simulated as \Bz\ decays to \aunobpi\
with \aunorhopi.

In the reconstruction of these decays we require
$0.87<m_{\aunob}<1.8$ \gev\
and  $0.51<m_{\rho}<1.1$ \gev.
We impose  several PID requirements to ensure the
identity of the signal pions. 
A $B$ candidate is characterized kinematically by the energy-substituted 
mass $\mes = \sqrt{(s/2 + \pvec_0\cdot \pvec_B)^2/E_0^2 - \pvec_B^2}$ and
energy difference $\DE = E_B^*-\half\sqrt{s}$, where the subscripts $0$ and
$B$ refer to the initial \UfourS\ and to the $B$ candidate in the
laboratory frame, respectively, 
and the asterisk denotes the CM frame. The resolutions in \mes\ and 
in \DE\ are  about 3.0  \mev\ and  20 \mev\ respectively.
We require $|\DE|\le0.1$ GeV and $5.25\le\mes\le5.29\ \gev$. 

To reject continuum background, we use
the angle $\theta_T$ between the thrust axis of the $B$ candidate and
that of the rest of the tracks and neutral clusters in the event, calculated in
the CM  frame.  We require $|\cos{\theta_T}|<0.65$
To suppress further combinatorial background we require that  the absolute value of 
the cosine  of the angle between the direction of the
$\pi$ meson from  $\aunob \ra \rho \pi$   with respect to the flight direction of the $B$ 
in the \aunob\ meson rest frame is required to be less than $0.85$.  
We discriminate further against \qqbar\ background with a
Fisher discriminant \xf\ that combines several variables~\cite{a1pi}.   

We use MC simulations of \BzBzb\ and \BpBm\ decays to look for \BB\ backgrounds. 
Neutral and charged $D$ mesons may contribute to background  through particle mis-identification or 
mis-reconstruction. We remove any combinations of the decay products,
including possible additional $\pi^0$, with  invariant mass consistent with  nominal mass values for 
$D^{\pm}\ra K^{\mp} \pi^{\pm}\pi^{\pm}$ or $\KS\ \pi^{\pm}$ 
and  $D^0\ra K^{\mp} \pi^{\pm}$  or $K^{\mp}\pi^{\pm}\pi^0$. 
The decay mode \Nbtoadueppim\   has the same 
final-state particles as the signal. We improve the discrimination against   
this decay with an angular variable \angular\ , defined as  the cosine
of the angle between the normal to the plane of the $3\pi$ resonance
and the flight direction of the primary pion from $B$ meson evaluated in the $3\pi$
resonance rest frame. We require $|\angular|< 0.62$.

\section{The Maximum Likelihood Fit}
We obtain the \CP  parameters and signal yield from an unbinned extended  maximum 
likelihood (ML) fit with the input observables \DE, \mes, \xf,
$m_{\aunob}$, \angular, and \deltat.  
We have six fit components in the likelihood: signal, charm and charmless \BB\ background, 
 \Nbtoadueppim, continuum \qqbar\ background, and non-resonant $\rho \pi \pi $. 
The flavor-tagging algorithm uses six mutually exclusive  categories \cite{Vincenzo}. 

The total probability density function (PDF) for the component $j$ and 
tagging category  $c$ in the event $i$,  ${\cal P}_{j,c}^i$,  is written as a product 
of the PDFs of the discriminating variables used in the fit.
The factored form of the PDF is a good approximation
since linear correlations among observables are below 10\%.
The systematic uncertainty from residual correlations is taken into
account in the fit bias.
We write the extended likelihood function for all events as
\begin{equation}
{{\cal L}} =  \prod_c  \exp{(-n_c)} \prod_i^{N_c} \left[ \sum_{j}n_j
f_{j,c}  {\cal P}_{j,c}^i  \right]\,,
\end{equation}
where $n_j$ is the yield of events of component $j$,  $f_{j,c}$ is the fraction of events
of component $j$ for each category $c$, 
$n_c = \sum_j f_{j,c}n_j $  is the number of events found by the fitter for category $c$, and $N_c$ is 
the number of events of category $c$ in the sample.  
We fix $f_{j,c}$  to $f_{\bflav,c}$, the values measured with a
large
sample of fully reconstructed $B^0$ decays into flavor eigenstates
(\bflav\ sample) \cite{Resol}, for the  signal, $\rho\pi\pi$, and  \Nbtoadueppim\
fit components.  We fix $f_{j,c}$ to values obtained with MC events for 
the charmless and charm fit components and allow it to vary for  the \qqbar\ component.

We test and calibrate the fitting procedure by applying it to
ensembles of simulated \qqbar\ experiments drawn from the PDF, into which
we have embedded the expected number of signal, charmless, 
 \Nbtoadueppim, charm, and  $\rho \pi \pi $ 
events randomly extracted from the fully simulated MC samples. The
measured quantities  $\Sa1pi$, $\Ca1pi$,
$\dSa1pi$, $\dCa1pi$, and $\Acpapi$
have been corrected for the fit biases and a systematic uncertainty
equal to half of the bias found in MC simulations is assigned  on the final results. 

In the fit there are 35 free parameters, including $\Sa1pi$, $\Ca1pi$,
$\dSa1pi$, $\dCa1pi$, the  charge asymmetries  for signal and 
continuum background, five yields, 
the signal $\aunob$ width, eleven parameters determining the shape of the combinatorial 
background, and 12 tagging efficiencies for the continuum.

\section{Results}
 The maximum likelihood fit to a sample of 29300 events results in a signal yield of
$608 \pm 53$, of which $461 \pm 46$ have their flavor identified.

\begin{figure}[h]
\resizebox{\columnwidth}{!}{
\begin{tabular}{cc}
\includegraphics[scale=0.25]{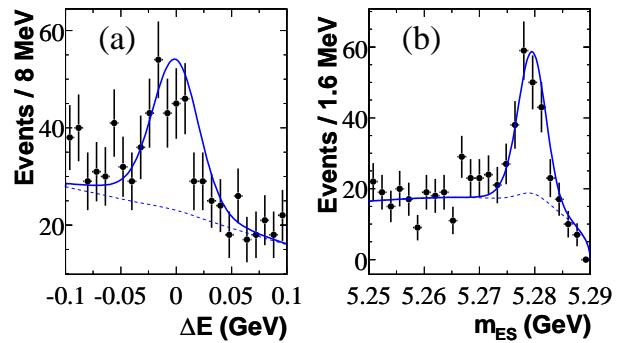} 
\end{tabular}
}
\vspace*{-0.5cm}
\caption{Projections of a) \DE, b)  \mes. 
Points represent on-resonance data, dotted lines 
the sum of all backgrounds, and solid lines the full fit
function.}
  \label{fig:ProjMesDE}
\end{figure}

Figure~\ref{fig:ProjMesDE} shows distributions of $\mes$ and $\DE$, 
enhanced in signal content by requirements on the signal-to-continuum 
likelihood ratios using all discriminating variables other than the
one plotted.

\begin{figure}[h]
\includegraphics[scale=0.35]{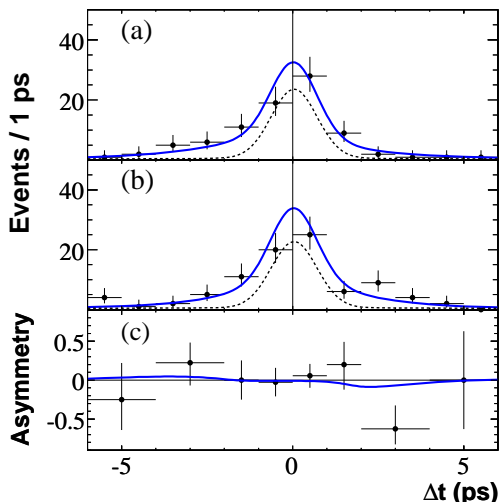} 
\vspace*{-0.5cm}
 \caption{Projections onto $\Delta t$  
 of the data (points) for a) \Bz\ 
and b) \Bzb\  tags, showing the fit function (solid 
line), and the background
function (dotted  line), and c) the asymmetry between \Bz\ and \Bzb\ 
tags.}
  \label{fig:DeltaTProj}
\end{figure}

Figure~\ref{fig:DeltaTProj} gives the $\Delta t$
projections and asymmetry for flavor tagged events selected as 
for Fig.~\ref{fig:ProjMesDE}.

We have studied  systematic uncertainties arising  from several
sources: variation of the signal PDF shape parameters within their
errors; modeling of the signal \deltat\ distribution;   tagging efficiency 
and mistag rates  determined from
the \bflav\ sample; uncertainties in 
$\deltamd$ and $\tau$ \cite{PDG2006}; uncertainty in the fit bias; 
uncertainty due to \CP\ violation present in the \BB\ background, the \Nadueppim\
\CP\ violation; uncertainty due to the interference 
between $B^0  \ra a_1^{\pm} \pi^{\mp}$ and other $4\pi$ final states;
doubly-Cabibbo-suppressed (DCS)  $b \ra \bar{u} c
\bar{d}$ amplitude for some tag-side $B$ decays~\cite{Long}; SVT
alignment; and the particle identification algorithm. We allow for a \CP asymmetry 
up to 20\% in  $B$ decays to charmless final states, and up to 
50\% in $B$ decays to ${a_2(1320)} \pi$. The total systematic error (\%) on the fit parameters 
$\Sa1pi\ $ , $\Ca1pi\ $ , $\dSa1pi\ $ , $\dCa1pi\ $, and $\Acpapi\ $ are $7.0$, $8.5$, $6.4$, $7.1$, and $1.6$
respectively. 
 
We measure
$\Sa1pi   = 0.37 \pm0.21 \pm 0.07$, $\dSa1pi = -0.14 \pm0.21 \pm0.06 $, 
$\Ca1pi = -0.10\pm 0.15\pm0.09$, $\dCa1pi= 0.26 \pm 0.15\pm
0.07$, $\Acpapi   = -0.07 \pm 0.07 \pm0.02$. Linear correlations
between these fit parameters are small.

Using the measured  fit parameters in formula~\ref{eq:alfaeff}, we extract the angle $\alpha_{\rm eff}$ 
and  one of the four solutions,
$\alpha_{\rm eff} = 78.6^{\circ} \pm 7.3^{\circ}$, is compatible with
the result of SM-based fits. Using the published branching
fraction \cite{a1pi} and adding statistical
and systematic errors in quadrature, we derive from relation~\ref{eq:thTimee}
the following values for the flavor-charge branching
fractions (in units of  $10^{-6}$): ${\cal B}(\Bz \ra a_1^+ \pi^-)=17.9\pm4.8$,
${\cal B}(\Bz \ra a_1^- \pi^+)=11.4\pm4.7$,
${\cal B}(\Bzb \ra a_1^+ \pi^-)=13.0\pm4.3$,
and ${\cal B}(\Bzb \ra a_1^- \pi^+)=24.2\pm5.8$.
The average of the branching fractions in the decays   $\Bz \ra a_1^+ \pi^-$ and  
$\Bzb \ra a_1^- \pi^+$, where the \aunob\ meson is emitted by the W boson, is larger than 
that in the decays  $\Bz \ra a_1^- \pi^+$ and  $\Bzb \ra a_1^+ \pi^-$, where the \aunob\ 
meson originates from the spectator interaction. This behaviour is in agreement 
with expectations based on form factor arguments.   

\section{Summary }
 In summary, we measure in the $\Bz(\Bzb) \ra  a_1^{\pm}\pi^{\mp}$  decays
 the charge asymmetry
$\Acpapi = -0.07 \pm 0.07 \pm 0.02$, 
 the mixing-induced \CP violation parameter  $\Sa1pi = 0.37 \pm 0.21\pm 0.07$, the direct \CP violation 
parameter $\Ca1pi = -0.10 \pm 0.15\pm 0.09$, and the parameters $\dCa1pi = 0.26 \pm 0.15\pm 0.07$ 
and $\dSa1pi = -0.14 \pm 0.21 \pm 0.06$. From these measured quantities we extract  the angle 
$\alpha_{\rm eff} = 78.6^{\circ} \pm 7.3^{\circ}$ and 
the following values for the flavor-charge branching
fractions (in units of  $10^{-6}$): ${\cal B}(B^0\ra a_1^+ \pi^-)=17.9\pm4.8$,
${\cal B}(B^0\ra a_1^- \pi^+)=11.4\pm4.7$,
${\cal B}(\Bbar^0\ra a_1^+ \pi^-)=13.0\pm4.3$,
and ${\cal B}(\Bbar^0\ra a_1^- \pi^+)=24.2\pm5.8$.
Once the  
measurements of branching fractions for SU(3)-related decays become
available, an upper  bound on $\Delta \alpha$ will provide a  constraint on the angle $\alpha$.
\\
\begin{acknowledgments}
 I thank Vincenzo Lombardo for helpful discussions.
\end{acknowledgments}

\end{document}